\documentclass{sf2a-conf2023}
\usepackage{graphicx}
\usepackage{hyperref}
\usepackage[]{natbib}  
\usepackage{epstopdf}

\def\BibTeX{{\rm B\kern-.05em{\sc i\kern-.025em b}\kern-.08em
    T\kern-.1667em\lower.7ex\hbox{E}\kern-.125emX}}
\bibpunct{(}{)}{;}{a}{}{,}  

\begin{document}

\TitreGlobal{SF2A 2023}


\title{STATUS OF WOMEN IN ASTRONOMY: A Need for Advancing Inclusivity and Equal Opportunities}

\runningtitle{Status of women in Astronomy}

\author{Mamta Pandey-Pommier$^{1,}$}\address{CNRS/Laboratoire Univers et Particules de Montpellier, Université de  Montpellier LUPM CC 072 - Place Eugène Bataillon 34095 Montpellier Cedex 5, France}
\address{University Catholic of Lyon, 10, place des Archives  69288 Lyon Cedex 02, France}
\author{Arianna Piccialli}\address{Royal Belgian Institute for Space Aeronomy
Planetary Aeronomy
Ringlaan 3 Av. Circulaire
1180 Brussels
Belgium}
\author{Belinda J. Wilkes$^{4,}$}\address{Center for Astrophysics Harvard $\&$ Smithsonian
SAO
Cambridge MA 02138-1516
Massachusetts (MA)
United States}
\address{Royal Society Wolfson Visiting Fellow, 
School of Physics, University of Bristol, 
United Kingdom}
\author{Priya Hasan}\address{Dept of Physics, Maulana Azad National Urdu University, Hyderabad, India 500034
}
\author{Santiago VargasDominguez}\address{Universidad Nacional de Colombia
Observatorio Astronómico
Av. Carrera 30, 45-03 Ciudad Universitaria
Bogota
Colombia}
\author{Alshaimaa Saad Hassanin}\address{Cairo University- Faculty of Science
Astronomy, Space Science $\&$ Meteorology Department
Cairo University-Faculty of Science- Astronomy Dep
Giza 12613
Egypt}
\author{Daniela Lazzaro}\address{Observatorio Nacional
Coordenação de Pesquisas em Astronomia e Astrofisica
R. Gal. Jose Cristino 77
São Cristóvão
Rio de Janeiro, RJ 20921 400
Rio de Janeiro (RJ)
Brazil}
\author{Claudia D. P. Lagos}\address{University of Western Australia
International Centre for Radio Astronomy Research
7 Fairway
Crawley 6009
Western Australia (WA)
Australia}
\author{Josefa Masegosa}\address{Inst de Astrofisica de Andalucía-CSIC
Extragalactic Astronomy
C/ Glorieta de la Astronomía
18008 Granada
Spain}
\author{Lili Yang}\address{School of Physics and Astronomy, Sun Yat-Sen University, No. 2 Daxue Road, 519082, Zhuhai China}
\author{David Valls-Gabaud}\address{Observatoire de Paris
LERMA
61 Avenue de l'Observatoire
75014 Paris
France
}
\author{John Leibacher$^{14,}$}\address{Lunar and Planetary Laboratory, University of 
Arizona (AZ), Tucson
85721}
\address{Institut d'Astrophysique Spatiale, CNRS, Universit\'e Paris-Saclay, UMR 8617, 91405 Orsay Cedex, France}
\author{Dara J. Norman}\address{NOIRLab
950 North Cherry Ave
Tucson AZ 85719
Arizona (AZ)
United States}
\author{Jolanta Nastula}\address{Polish Academy of Sciences
Space Research Center
Bartycka 18 A
00-716 Warsaw
Poland}
\author{Aya Bamba}\address{The University of Tokyo
Department of Physics
Hongo 7-3-1
Bunkyo-ku, Tokyo 113-1133
Kanto
Japan}



\setcounter{page}{237}


\maketitle


\begin{abstract}
Women in the Astronomy and STEM fields face systemic inequalities throughout their careers. Raising awareness, supported by detailed statistical data, represents the initial step toward closely monitoring hurdles in career progress and addressing underlying barriers to workplace equality. This, in turn, contributes to rectifying gender imbalances in STEM careers.  
The International Astronomical Union Women in Astronomy (IAU WiA) working group, a part of the IAU Executive Committee, is dedicated to increasing awareness of the status of women in Astronomy and supporting the aspirations of female astronomers globally. Its mission includes taking concrete actions to advance equal opportunities for both women and men in the field of astronomy. In August 2021, the IAU WiA Working Group established a new organizing committee, unveiling a comprehensive four-point plan. This plan aims to strengthen various aspects of the group's mission, encompassing:
 (i) Awareness Sustainability: Achieved through surveys and data collection, (ii) Training and Skill Building: Focused on professional development, (iii) Fundraising: To support key initiatives, and (iv) Communication: Dissemination of results through conferences, WG Magazines, newsletters, and more. This publication provides an overview of focused surveys that illuminate the factors influencing the careers of women in Astronomy, with a particular focus on the careers of mothers. It highlights the lack of inclusive policies, equal opportunities, and funding support for women researchers in the field. Finally, we summarize the specific initiatives undertaken by the IAU WiA Working Group to advance inclusivity and equal opportunities in Astronomy.
\end{abstract}



\newpage
\section{Introduction}
Worldwide, women researchers in the field of Astronomy face significant underrepresentation. Factors contributing to this situation include a lack of strong support for job applications, fewer opportunities, reduced participation in peer reviews for journals and funding proposals, fewer invitations to give colloquia compared to their counterparts at prestigious universities and in leadership roles on decision-making committees, caregiving, and family responsibilities \citep{Piccialli2020}. To address the gender imbalances at every career stage and promote international efforts to support women's careers in astronomy, the IAU Executive Committee (EC) established the Working Group for Women in Astronomy (IAU WiA) in 2003 \citep{Cesarsky2010}. Over the past two decades, the WiA WG has organized various activities and events at IAU General Assemblies, including Women's Luncheons, Students Mentoring Sessions, and Special Talks, to raise awareness about the lack of inclusivity and equal opportunity that women astronomers encounter in this field \citep{Primas2018, Norman2021}. 
Despite these efforts, the representation of women in the astronomy community has only marginally increased, and significant gender disparities persist in many parts of the world. To better understand the current status of women in astronomy, the acting IAU WiA WG conducted statistics of current IAU members data available at \url{https://www.iau.org/public/themes/member\_statistics/}. The aim was to gather statistical information and show the current status of the representation of women in the field of astronomy worldwide \citep{Pandey-Pommier2021}. 
\begin{figure}[!h]
  \includegraphics[width=0.66\textwidth, angle=270]{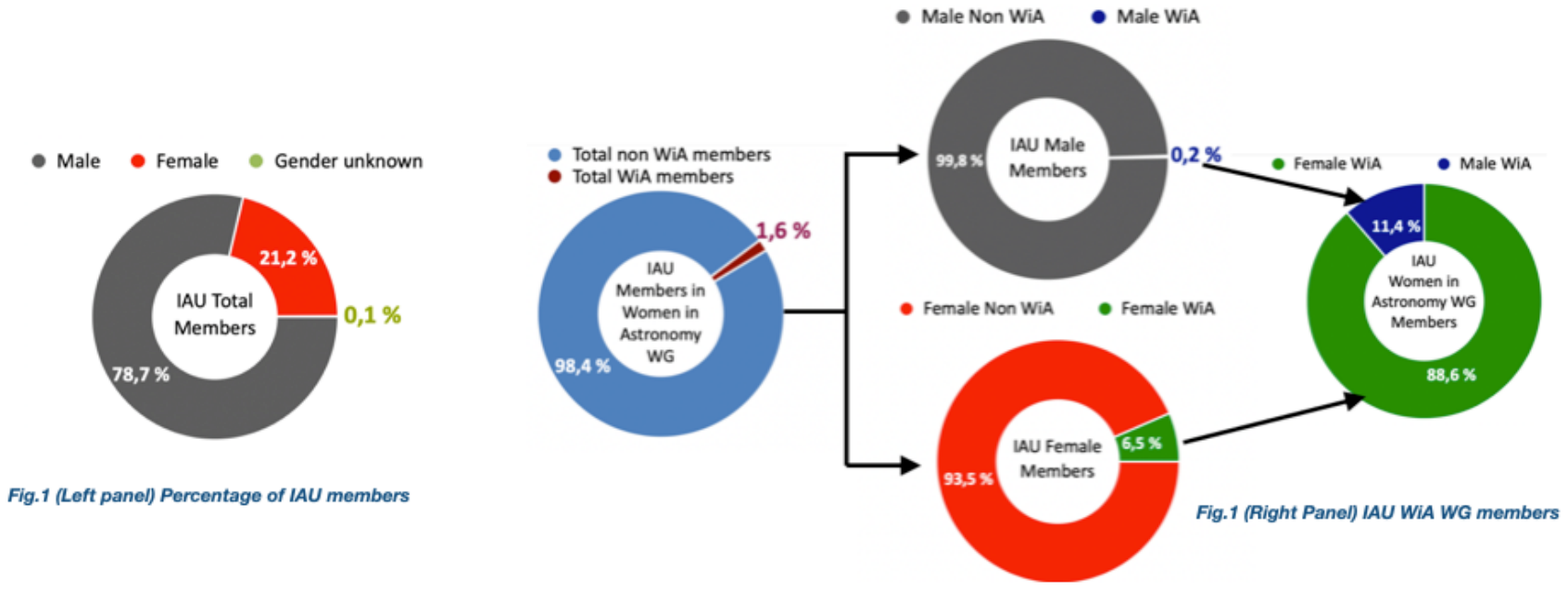}
  \caption{IAU members and participants of WiA WG, adapted from \citep{Pandey-Pommier2021}}
\label{figure}
\end{figure}

The statistical analysis results shown in Figure \ref{figure} revealed that the IAU has less than $<21\%$ female membership among its 12,000 total members, with fewer than $<1.6\%$ actively participating in the WiA WG. 
Furthermore, only 29 countries have women actively participating in research activities, comprising $\leq 21\%$ of the total researchers in each country. Among these 29 countries, 18 are located in the European Union, while the remaining 11 are distributed across the Americas (4), Asia(4), Africa(1), Oceania(1), and the Middle East(1). These findings underscore the minimal participation of the overall astronomy community members to achieve gender balance within the field as well as the low global representation of women in astronomy, leading to a biased and discriminatory working environment. 

\section{Survey data and observation}
To identify the critical factors influencing the underrepresentation and careers of women in Astronomy, raise awareness about the current working conditions for women astronomers, and emphasize the imperative need for global efforts to achieve inclusivity and gender balance in this field, M. Pommier conducted a comprehensive survey titled 'Working Conditions for Women in Astronomy', spanning over three years, beginning in 2019.  This survey aimed to identify various factors, their proportions, and their impact on the representation of women in Astronomy \citep{Pandey-Pommier2021}. In the present paper, we delve into the results obtained from this survey. The sections are organized as follows: Section 3.1 presents the findings on factors affecting the career growth of women astronomers, followed by Section 3.2, which discusses the 4-point action plan initiated by the IAU WiA WG to advance skill development and equal opportunity possibilities. Finally, in Section 4, we engage in a discussion about the changes in policies needed to support the career development of women in astronomy.

 \section{Results}
 A total of 750 participants from 58 countries took part in this survey, with more than $60\%$ of the respondents hailing from EU countries, followed by Americas($\sim 19\%$), Oceania($\sim 13\%$), Asia($\sim 4.5\%$), Africa($\sim 2.5\%$), and the Middle East($\sim 1\%$).
  \subsection{\bf Factors affecting the career of women in Astronomy}
\begin{figure}[h]
  \includegraphics[width=0.75\textwidth, angle=270]{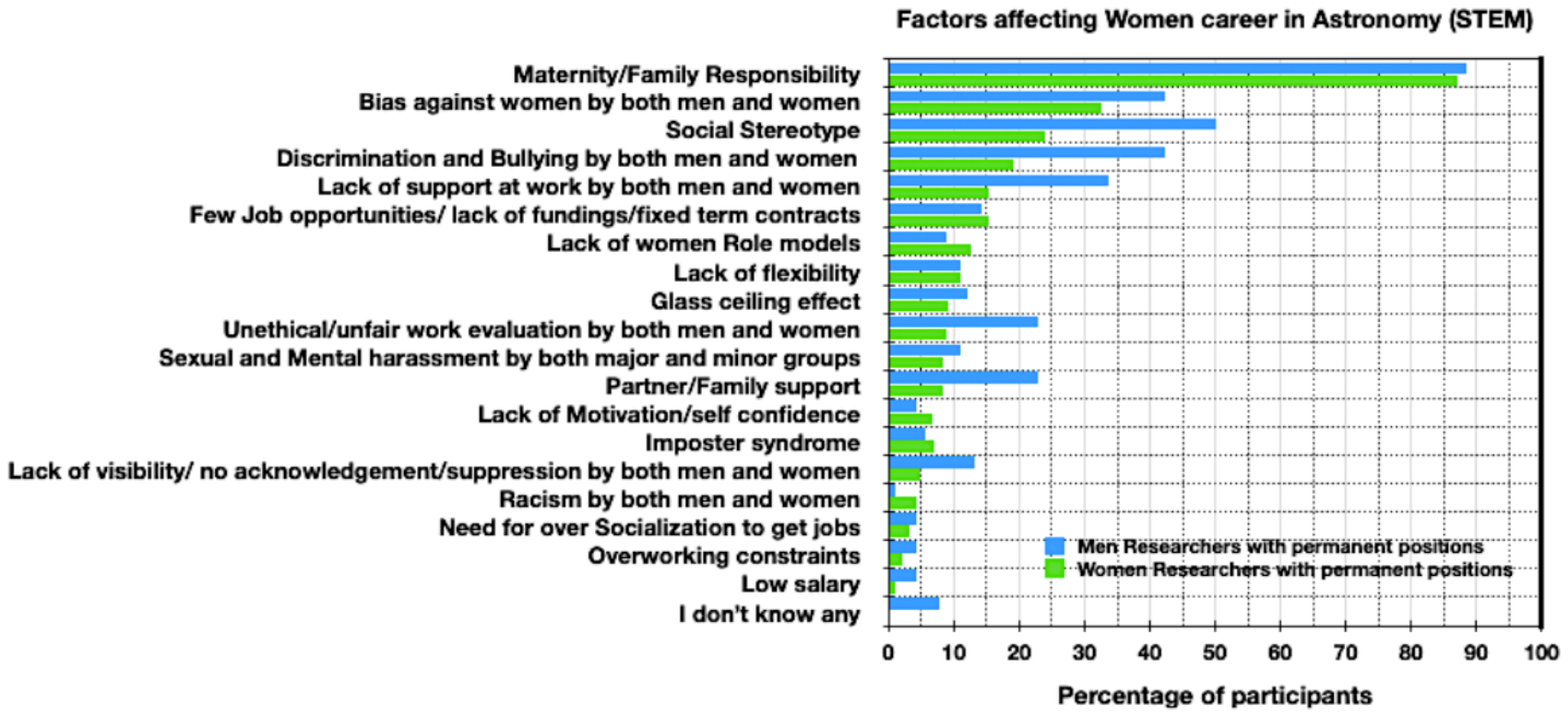}
  \caption{Survey Results on factors affecting Women's Career in Astronomy and STEM fields, adapted from \citep{Pandey-Pommier2021}}
\label{figure2}
\end{figure}

  Several factors have a significant impact on the careers of women in Astronomy, as illustrated in Figure \ref{figure2} with the most crucial:
  \begin{itemize}
    \item 
    Family and Caregiving Responsibilities: related to caring for children and elderly family members.
\item 
Bias Against Women Applicants, Especially Mothers: Discriminatory biases such as underpaid/unpaid contracts can affect the careers of women, particularly those who are mothers, during the application process.
\item 
Social Stereotypes: Gender, marital status, race, and origin can lead to harmful social stereotypes that hinder career progression.
\item 
Discrimination and Bullying: Discriminatory and bullying behaviors, both from older and younger colleagues, create hostile work environments.
\item 
Lack of Funds, Role Models, and Support for Job Applications: Women applicants, especially mothers, often face challenges in securing necessary funding and workplace support leading to extended years of fixed-term positions, that may be an unethical working condition 
\item 
Unethical Work Evaluation: Unfair work evaluations may result in projects of non-permanent female members systematically being taken over by permanent members or given to younger male colleagues without consent.
\item 
Lack of Family Support, Especially for Mothers: A lack of family support, particularly for mothers, can impact their ability to balance work and caregiving responsibilities.
\item 
Sexual and Mental Harassment: Both older and younger colleagues can engage in sexual and mental harassment, further hindering career advancement.
  \end{itemize} 
  
The findings of this survey have identified eight crucial factors that significantly impact the careers of women astronomers at every stage, with all of these factors predominantly affecting the careers of mothers in this field while mostly 3 (Sexual and Mental Harassment, Bullying, and Unethical work evaluation) affect the careers of all women in astronomy. This leads to a {\bf systematic leaky pipeline effect in the career of women in Astronomy leading to underrepresentation and gender imbalance in this field}.  It is thus imperative to urge policymakers to formulate inclusive policies to address these long-standing issues faced by women, particularly the challenges encountered by mothers, in academia. As of now the situation of mothers is poorly/inadequately addressed in hiring and evaluation committees policies. Consequently, this study underscores the urgent need for global efforts and increased awareness to achieve gender balance in the field of Astronomy and to collaborate with large international organizations dedicated to supporting women in STEM. \citep{Pandey-Pommier2021}.

  \subsection{\bf Activities initiated by the IAU WiA WG}
  Taking into account the marginalized and reduced job prospects for women in Astronomy, the Organizing Committee of the IAU WiA WG initiated a four-point action plan to advance the careers of women in Astronomy, \textbf{(ref. IAU WiA WG newsletters)}- \url{https://sites.google.com/view/iau-women-in-astronomy/home/iau-wia-ensemble-magazine-newsletter?authuser=2}, by creating:
\begin{itemize}
    \item 
 Awareness and Sustainability- This involves conducting surveys on the working conditions of women researchers, collecting statistical data from around the world, and highlighting gender balance issues within the IAU and the global astronomical community. Survey results are regularly featured in WG ENSEMBLE Magazine and annual reports, fostering a self-supporting and sustainable community\citep{Pandey-Pommier2022}.
 \item 
 Training and Skill Building- The focus here is on developing resources, gathering materials from other IAU offices, and conducting regular training programs to enhance the skills and professional development of female researchers worldwide\citep{Hasan2022}.
 \item 
 Fundraising- This component aims to raise funds and offer financial support to women researchers at critical stages in their careers. 
   \item 
 Communication and Dissemination- This involves promoting the participation of women researchers worldwide by establishing communication channels on social media and organizing regular seminar series. These seminars feature experts, both permanent and non-permanent female and male members (supporting careers of women in Astronomy), from a geographically diverse team with varying levels of experience \citep{Pandey-Pommier2022}.
\end{itemize}
We have received global appreciation and support for these initiatives and ongoing activities from both permanent and non-permanent members of the worldwide community. Nevertheless, securing long-term funding is essential to sustain and further our efforts in promoting gender balance and ensuring healthy working conditions in the fields of astronomy and STEM.

\section{Discussion and Future Prospects on Inclusivity and Equal Opportunities for Women in Astronomy}
This study underscores the current working conditions of women in Astronomy worldwide, which are affected by a lack of inclusive and equal opportunity work environment. The primary findings of our research are as follows:
\begin{itemize}
\item Marginalized representation of women astronomers (especially mothers) at the global level within the IAU membership and individual countries.
\item Identification of eight major factors affecting the careers of women in Astronomy: maternity/family responsibilities, bias, stereotypes, discrimination/bullying, lack of funding, suppression of work, lack of support, and sexual/mental harassment. While all eight factors impact the careers of mothers, mainly three affect the careers of women without family and childcare responsibilities.
\item Recognition of the challenges in addressing the leaky pipeline effect within research fields, highlighting the absence of proper policies to date. There is an urgent need for comprehensive policies to combat this long-standing issue, which contributes to gender imbalance in the field.
\item A call for key policymakers, including funding agencies, universities/research organizations, and peer-reviewed journals, to implement inclusive and equal opportunity policies and criteria within their decision-making, hiring, and evaluation committees. These committees should comprise gender-balanced and diverse team members to comprehensively combat the leaky pipeline effect and other factors impacting women's careers in research. Particular attention needs to be paid to the situation of mothers who are working under reduced and non-inclusive opportunities or prolonged fixed-term contracts under unfair working conditions for several decades.
\end{itemize}

\begin{acknowledgements}
MP gratefully acknowledges the support of the Indo-French Centre for the Promotion of Advanced Research (Centre Franco-Indien
pour la Promotion de la Recherche Avancée) under project 6504-3.
\end{acknowledgements}

\bibliographystyle{aa}  
\bibliography{POMMIER_S09} 

\end{document}